\newif\ifpdf
\ifx\pdfoutput\undefined
\pdffalse 
\else
\pdfoutput=1 
\pdftrue \fi

\newif\iffinal
\finaltrue

\documentclass[reqno,twoside,final]{myjstatphys}
\iffinal\else\usepackage[notref,notcite]{showkeys}\fi
\usepackage{cite}
\usepackage{amsmath}
\usepackage{amsfonts}
\usepackage{amssymb}
\usepackage{graphicx}
\usepackage{graphics}
\usepackage{verbatim}

\input{ABC-I.aux}

\IfFileExists{epsf.def}{\input epsf.def}{\usepackage{epsf}}

\IfFileExists{myowntimes.sty}
	{\usepackage{myowntimes}}
	{\usepackage{times}
		\IfFileExists{mathrsfs.sty}
			{\usepackage{mathrsfs}}
			{\newcommand{\mathscr}{\mathcal}}}
\newenvironment{proofsect}[1]
{\par\normalfont\vskip0.3cm\noindent
{\hskip10.4mm\sffamily\slshape#1.}}
{\qed\vspace{0.15cm}}

\theoremstyle{jsp}

\newcommand{\textd}{\text{\rm d}}

\renewcommand{\AA}{\mathcal A}
\newcommand{\BB}{\mathcal B}

\newcommand{\RR}{\mathcal R}

\newcommand{\E}{\mathbb E}

\newcommand{\BbbP}{\mathbb P}

\newcommand{\R}{\mathbb R}

\newcommand{\Z}{\mathbb Z}

\newcommand{\scrM}{\mathscr{M}}

\newcommand{\scrQ}{\mathscr{Q}}

\newcommand{\scrS}{\mathscr{S}}

\newcommand{\scrW}{\mathscr{W}}

\newcommand{\twoeqref}[2]{(\ref{#1}--\ref{#2})}
\newcommand{\1}{{\text{\sf 1}}}

\newcommand{\sS}{{\small\textsl{S}}}
\newcommand{\barsS}{\,\overline{\!\small\textsl{S}}}

\newcommand{\ccdot}{\mkern1mu\cdot\mkern1.3mu}

\newcommand{\Jc}{J_{\text{\rm c}}}

\newcommand{\mstar}{m_\star}

\newcommand{\xit}{\xi_{\text{\rm t}}}
\newcommand{\txit}{\tilde{\xi}_{\text{\rm t}}}
\newcommand{\txiu}{\tilde{\xi}_{\text{\rm u}}}
\newcommand{\tb}{\tilde{b}}
\newcommand{\oE}{\widehat{E}}
\newcommand{\mT}{m_{\text{\rm T}}}

\newcommand{\bn}{\boldsymbol n}

\begin{document}

\title[Colligative properties of solutions]{Colligative properties of
solutions:\\*[2mm]II.~Vanishing concentrations}

\author{Kenneth~S.~Alexander,\footnotemark[1] Marek~Biskup,\footnotemark[2] and Lincoln~Chayes\footnotemark[2]}

\renewcommand{\thefootnote}{}
\footnotetext{\hglue-1.9em$\copyright$\,2004 by K.S.~Alexander, M.~Biskup and
L.~Chayes.  Reproduction, by any means, of the entire article for
non-commercial purposes is permitted  without~charge.}
\renewcommand{\thefootnote}{\arabic{footnote}}

\footnotetext[1]{Department of Mathematics, USC, Los  Angeles, California, USA}
\footnotetext[2]{Department of Mathematics, UCLA, Los Angeles, California, USA.}

\runningauthor{Alexander, Biskup and Chayes}

\begin{abstract}
We continue our study of colligative
properties of solutions initiated in ref.~\cite{ABC}. We focus on the situations where, in a system of linear size~$L$, the concentration and the chemical potential scale like~$c=\xi/L$ and~$h=b/L$, respectively. We find that there exists a critical value~$\xit$ such that no phase separation occurs for~$\xi\le\xit$ while, for~$\xi>\xit$, the two phases of the solvent coexist
for an interval of values of~$b$. Moreover, phase separation begins abruptly in the sense that a
macroscopic fraction of the system suddenly freezes (or melts)
forming a crystal (or droplet) of the complementary phase when~$b$ reaches
a critical value.  For certain values of system parameters, under ``frozen''
boundary conditions, phase separation also ends abruptly in the sense that the equilibrium droplet grows continuously with increasing~$b$ and then suddenly jumps in size to subsume the entire system. Our findings indicate that the onset of freezing-point depression is in fact a surface~phenomenon.
\end{abstract}

\section{Introduction}
\subsection{Overview}
\label{sec1.1}\noindent In a previous paper (ref.~\cite{ABC},
henceforth referred
to as Part~I) we defined a model of non-volatile solutions and studied its
behavior under the conditions when the solvent undergoes a liquid-solid phase
transition. A particular example of interest is the solution of salt
in water at
temperatures near the freezing point. In accord with Part~I we will
refer to the
solute as salt and to the two phases of solvent as ice and liquid water.

After some reformulation the model is reduced to the Ising model coupled to an
extra collection of variables representing the salt. The (formal)
Hamiltonian is given by
\begin{equation}
\label{1.4}
\beta\mathscr{H}= -J\sum_{\langle x,y\rangle}\sigma_x\sigma_y -h\sum_x\sigma_x
+\kappa\sum_x\sS_x\frac{1-\sigma_x}2.
\end{equation}
Here we are confined to the sites of the hypercubic
lattice~$\Z^d$
with~$d\ge2$, the variable~$\sigma_x\in\{+1,-1\}$ marks the presence of liquid
water ($\sigma_x=1$) and ice ($\sigma_x=-1$) at site~$x$,
while~$\sS_x\in\{0,1\}$
distinguishes whether salt is present ($\sS_x=1$) or absent ($\sS_x=0$) at~$x$.
The coupling between the~$\sigma$'s is ferromagnetic ($J>0$), the coupling
between the~$\sigma$'s and the~$\sS$'s favors salt in liquid water,
i.e.,~$\kappa>0$---this reflects the fact that there is an energetic
penalty for salt inserted into the crystal structure of ice.

A statistical ensemble of direct physical---and
mathematical---relevance is that
with fluctuating magnetization (grand canonical spin variables) and a
fixed amount of salt (canonical salt variables). The principal
parameters
of the system are thus the salt concentration~$c$ and the external
field~$h$. As
was shown in Part~I for this setup, there is a non-trivial region in
the~$(c,h)$-plane where phase separation occurs on a macroscopic scale.
Specifically, for~$(c,h)$ in this region, a droplet which takes a non-trivial 
(i.e., non-zero and non-one) fraction of the
entire volume appears in the system.
(For ``liquid'' boundary conditions, the droplet is actually an
ice crystal.) In ``magnetic'' terms, for each~$h$ there is a unique
value of the magnetization which is achieved by keeping part of the
system in the liquid, i.e., the plus Ising state, and part in the solid, i.e., the
minus Ising state. This is in sharp contrast to what happens in the unperturbed
Ising model where a single value of~$h$ (namely,~$h=0$) corresponds to a whole
\emph{interval} of possible magnetizations.

The main objective of the present paper is to investigate the limit of
infinitesimal salt concentrations. We will take this to mean the
following: In a
system of linear size~$L$ we will consider the above ``mixed'' ensemble with
concentration~$c$ and external field~$h$ scaling to zero as the size of the
system,~$L$, tends to infinity. The goal is to describe the
asymptotic properties
of the typical spin configurations, particularly with regards to the
formation of
droplets. The salt marginal will now be of no interest because salt particles are
so sparse that any local observable will eventually report that there is no
salt at all.

\begin{figure}[t]
\refstepcounter{obrazek}
\label{fig2}
\vspace{.2in}
\ifpdf
\centerline{\includegraphics[width=5.3in]{scalediag.pdf}}
\else
\centerline{\includegraphics[width=5.3in]{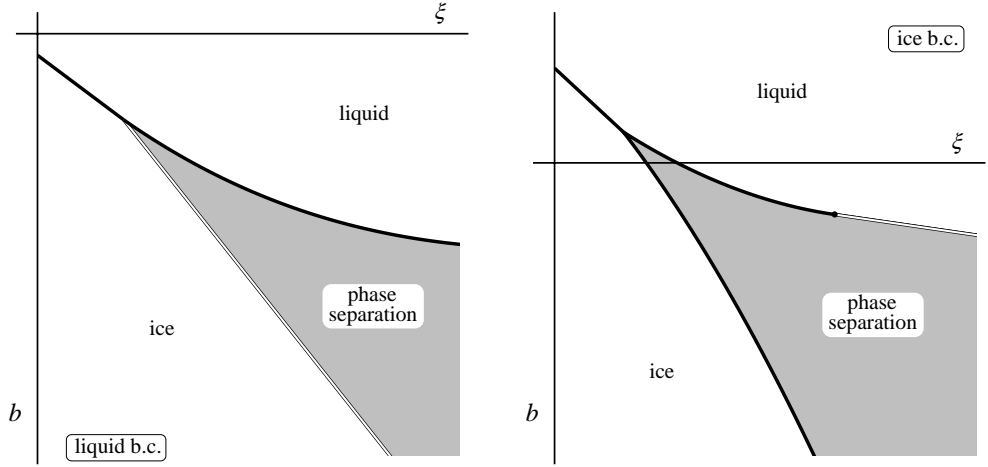}}
\fi
\caption{ 
The phase diagram of the ice-water system with Hamiltonian~\eqref{1.4} and fixed salt concentration~$c$ in a Wulff-shaped vessel of linear size~$L$. The left plot
corresponds to the system with plus boundary conditions,
concentration $c=\xi/L$
and field parameter $h=b/L$, the plot on the right depicts the situation for minus boundary conditions. It is noted that
as~$\xi$ ranges in~$(0,\infty)$ with~$b$ fixed, three distinct modes
of behavior emerge, in the~$L\to\infty$ limit, depending on the value of~$b$.
The thick black lines mark the phase boundaries where a droplet transition occurs; on the white lines the fraction of liquid (or solid) in the system changes continuously.}
\end{figure}

The main conclusions of this work are summarized as follows. First,
in a regular
system of volume~$V=L^d$ of characteristic dimension~$L$, the scaling for both
the salt concentration and external field is~$L^{-1}$. In
particular, we should write~$h=bL^{-1}$ and~$c=\xi L^{-1}$. Second, considering such a system with boundary condition favoring the liquid state and with~$h$
and~$c$ enjoying the abovementioned scalings, one of three things
will happen as~$\xi$ sweeps from~$0$ to infinity:
\begin{enumerate}
\item[(1)] 
If~$b$ is sufficiently small negative, the system is always in the
liquid state.
\item[(2)] 
If~$b$ is of intermediate (negative) values, there is a
transition, at some~$\xi(b)$ from the ice state to the liquid state.
\item[(3)] 
Most dramatically, for larger (negative) values of~$b$, there is a
region---parametrized by~$\xi_1(b)<\xi<\xi_2(b)$---where (macroscopic)
phase separation occurs. Specifically, the system holds a large crystalline chunk of
ice, whose volume fraction varies from unity to some \emph{positive} amount as~$\xi$ varies
from~$\xi_1(b)$ to~$\xi_2(b)$. At~$\xi=\xi_2(b)$, all of the remaining ice suddenly melts.
\end{enumerate}
We obtain analogous results when the boundary conditions favor the ice state,
with the ice crystal replaced by a liquid ``brine pocket.'' However, here a
new phenomenon occurs: For certain choices of system parameters, the (growing) volume
fraction occupied by the brine pocket remains bounded away from one as~$\xi$ increases from~$\xi_1(b)$ to~$\xi_2(b)$, and then jumps discontinuously to one at $\xi_2(b)$.
In particular, there are two droplet transitions, see~Fig.~\ref{fig2}.

Thus, we claim that the onset of freezing point depression is, in
fact, a \emph{surface} phenomenon. Indeed, for very weak solutions, 
the bulk behavior of the system is determined by a delicate balance between surface order deviations of the temperature and salt concentrations. In somewhat poetic terms, the predictions
of this work are that at the liquid-ice coexistence temperature it is possible to melt a substantial portion of the ice via a pinch of
salt whose size is only of the order~$V^{1-\frac1d}$. (However, we
make no claims as to how long one would have to wait in order to observe this phenomenon.)

\smallskip 
The remainder of this paper is organized as follows. In the next
section we reiterate the basic setup of our model and introduce some further
objects of relevance. The main results are stated in
Sections~\ref{sec2.1}-\ref{sec2.4}; the
corresponding proofs come in Section~\ref{sec3}. In order to keep the
section and
formula numbering independent of Part~I; we will prefix the
numbers from Part~I by~``I.''

\newcounter{obrazek}

\subsection{Basic objects}
\label{sec1.2}\noindent We begin by a quick reminder of the model; further
details and motivation are to be found in Part~I. Let~$\Lambda\subset\Z^d$ be a
finite set and let~$\partial\Lambda$ denote its (external) boundary. For each
$x\in\Lambda$, we introduce the water and salt variables,
$\sigma_x\in\{-1,+1\}$
and $\sS_x\in\{0,1\}$; on~$\partial\Lambda$ we will consider a fixed
configuration~$\sigma_{\partial\Lambda}\in\{-1,+1\}^{\partial\Lambda}$. The
finite-volume Hamiltonian is then a function of~$(\sigma_\Lambda,\sS_\Lambda)$
and the boundary condition~$\sigma_{\partial\Lambda}$ that takes the form
\begin{equation}
\beta\mathscr{H}_\Lambda(\sigma_\Lambda,\sS_\Lambda|\sigma_{\partial\Lambda})=
-J\!\!\!\sum_{\begin{subarray}{c}
\langle x,y\rangle\\ x\in\Lambda,\,y\in\Z^d
\end{subarray}}\!\!
\sigma_x\sigma_y
-h\sum_{x\in\Lambda}\sigma_x+\kappa\sum_{x\in\Lambda}\sS_x\frac{1-\sigma_x}2.
\end{equation}
Here, as usual,~$\langle x,y\rangle$ denotes a nearest-neighbor
pair on~$\Z^d$ and the parameters~$J$,~$\kappa$ and~$h$ represent the chemical
affinity of water to water, negative affinity of salt to ice and the
difference
of the chemical potentials for liquid-water and ice, respectively.

The \emph{a priori} probability distribution of the
pair~$(\sigma_\Lambda,\sS_\Lambda)$ takes
the usual Gibbs-Boltzmann
form~$P_\Lambda^{\sigma_{\partial\Lambda}}(\sigma_\Lambda,\sS_\Lambda)\propto
e^{-\beta
\mathscr{H}_\Lambda(\sigma_\Lambda,\sS_\Lambda|\sigma_{\partial\Lambda})}$.
For
reasons explained in Part~I, we will focus our attention on the ensemble with a
fixed total amount of salt. The relevant quantity is defined by
\begin{equation}
N_\Lambda=\sum_{x\in\Lambda}\sS_x.
\end{equation}
The main object of interest in this paper is then the
conditional
measure
\begin{equation}
P_\Lambda^{\sigma_{\partial\Lambda},c,h}(\cdot)=
P_\Lambda^{\sigma_{\partial\Lambda}}
\bigl(\,\cdot\,\big|N_\Lambda=\lfloor c|\Lambda|\rfloor\bigr),
\end{equation}
where~$|\Lambda|$ denotes the number of sites in~$\Lambda$. We
will mostly focus on the situations when~$\sigma_{\partial\Lambda}\equiv+1$
or~$\sigma_{\partial\Lambda}\equiv-1$, i.e., the plus or minus boundary
conditions. In these cases we denote the above measure
by~$P_\Lambda^{\pm,c,h}$,
respectively.

\smallskip
The surface nature of the macroscopic phase separation---namely, the
cases when the concentration scales like the inverse linear scale of the system---indicates that
the quantitative aspects of the analysis may depend sensitively on the
shape of the volume in which the model is studied. Thus, to keep this work
manageable, we will restrict our rigorous treatment of these cases to
volumes of
a particular shape in which the droplet cost is the same as in infinite volume.
The obvious advantage of this restriction is the possibility of explicit
calculations; the disadvantage is that the shape actually depends on
the value of
the coupling constant~$J$. Notwithstanding, we expect that all of our findings
are qualitatively correct even in rectangular volumes but that cannot be
guaranteed without a fair amount of extra work; see~\cite{SS2} for an example.

Let~$V\subset\R^d$ be a connected set with connected complement and
unit Lebesgue
volume. We will consider a sequence~$(V_L)$ of lattice volumes which are just
discretized blow-ups of~$V$ by scale factor~$L$:
\begin{equation}
\label{1.5} V_L=\{x\in\Z^d\colon x/L\in V\}.
\end{equation}
(The sequence of~$L\times\dots\times L$ boxes~$(\Lambda_L)$ from Part~I is
recovered by letting~$V=[0,1)^d$.) The particular ``shape''~$V$ for
which we will
prove the macroscopic phase separation coincides with that of an
equilibrium droplet---the \emph{Wulff-shaped volume}---which we will
define next.
We will stay rather succinct; details and proofs can be found in standard
literature on Wulff construction
(\cite{ACC,DKS,Bob+Tim,bigBCK,Bodineau,Cerf-Pisztora} or the
review~\cite{BIV}).
Readers familiar with these concepts may consider skipping the rest of this
section and passing directly to the statements of our main results.

\smallskip 
Consider the ferromagnetic Ising model at coupling~$J\ge0$ and zero
external field and let~$\BbbP_\Lambda^{\pm,J}$ denote the corresponding Gibbs
measure in finite volume~$\Lambda\subset\Z^d$ and plus/minus boundary
conditions.
As is well known, there exists a number~$\Jc=\Jc(d)$, with~$\Jc(1)=\infty$
and~$\Jc(d)\in(0,\infty)$ if~$d\ge2$, such that for every~$J>\Jc$ the
expectation
of any spin in~$\Lambda$ with respect to~$\BbbP_\Lambda^{\pm,J}$ is
bounded away
from zero uniformly in~$\Lambda\subset\Z^d$. The limiting value of this
expectation in the plus state---typically called the \emph{spontaneous
magnetization}---will be denoted by~$\mstar=\mstar(J)$. (Note that~$\mstar=0$
for~$J<\Jc$ while~$\mstar>0$ for~$J>\Jc$.)  

Next we will recall the basic setup
for the analysis of surface phenomena. For each unit vector~$\bn\in\R^d$, we
first define the surface free energy~$\tau_J(\bn)$ in direction~$\bn$. To this
end let us consider a rectangular box $V(N,M)\subset\R^d$ with
``square'' base of
side~$N$ and height~$M$ oriented such that~$\bn$ is orthogonal to the base. The
box is centered at the origin. We let~$Z^{+,J}_{N,M}$ denote the
Ising partition
function in~$V(N,M)\cap\Z^d$ with plus boundary conditions. We will
also consider
the inclined Dobrushin boundary condition which takes  value~$+1$ at the
sites~$x$ of the boundary of~$V(N,M)\cap\Z^d$ for which~$x\cdot\bn>0$
and~$-1$ at
the other sites. Denoting the corresponding partition function
by~$Z^{\pm,J,\bn}_{N,M}$, the surface free energy~$\tau_J(\bn)$ is
then defined by
\begin{equation}
\tau_J(\bn)=-\lim_{M\to\infty}\lim_{N\to\infty}\frac1{N^{d-1}}
\log\frac{Z^{\pm,J,\bn}_{N,M}}{Z^{+,J}_{N,M}}.
\end{equation}
The limit exists by subadditivity arguments. The quantity~$\tau_J(\bn)$ determines the cost of an interface orthogonal to vector~$\bn$.

As expected, as soon as~$J>\Jc$,
the function~$\bn\mapsto\tau_J(\bn)$ is uniformly positive~\cite{Lebowitz-Pfister}. 
In order to evaluate the cost of a curved interface,~$\tau_J(\bn)$ will have to
be integrated over the surface. Explicitly, we will let~$J>\Jc$ and, given a
bounded set~$V\subset\R^d$ with piecewise smooth boundary, we define the
\emph{Wulff functional}~$\scrW_J$ by the integral
\begin{equation}
\label{1.7}
\scrW_J(V)=\int_{\partial V}\tau_J(\bn)\,\textd A,
\end{equation}
where~$\textd A$ is the (Hausdorff) surface measure and~$\bn$ is
the position-dependent unit normal vector to the surface. The \emph{Wulff
shape}~$W$ is the unique minimizer (modulo translation) of~$V\mapsto\scrW(V)$
among bounded sets~$V\subset\R^d$ with piecewise smooth boundary and unit
Lebesgue volume. We let~$(W_L)$ denote the sequence of \emph{Wulff-shaped}
lattice volumes defined from~$V=W$ via \eqref{1.5}.

\section{Main results}
\label{sec2}\noindent 
We are now in a position to state and prove our main results. 
As indicated before, we will focus on the limit of
infinitesimal concentrations (and external fields) where~$c$ and~$h$ scale as the reciprocal of the
linear size of the system. Our results come in four theorems: In Theorem~\ref{thm3} we state the basic surface-order large-deviation principle. Theorems~\ref{thm4} and~\ref{thm4b} describe the minimizers of the requisite rate functions for liquid and ice boundary conditions, respectively. Finally, Theorem~\ref{thm5} provides some control of the spin marginal of the corresponding Gibbs measure.

\subsection{Large deviation principle for magnetization}
\label{sec2.1}\noindent 
The control of the regime under consideration involves
the surface-order large-deviation principle for the total magnetization in the
Ising model. In a finite set~$\Lambda\subset\Z^d$, the quantity under
considerations is given by
\begin{equation}
M_\Lambda=\sum_{x\in\Lambda}\sigma_x.
\end{equation}
Unfortunately, the rigorous results available at present for $d
\geq 3$ do not cover all of the cases to which our analysis might
apply. In order
to reduce the amount of necessary provisos in the statement of the theorems, we
will formulate the relevant properties as an assumption:

\bigskip\noindent
\textbf{Assumption~A\ }
\textsl{ Let~$d\ge2$ and let us consider a sequence of Wulff-shape
volumes~$W_L$.
Let $J>\Jc$ and recall that $\BbbP_{W_L}^{\pm,J}$ denotes the Gibbs
state of the
Ising model in~$W_L$, with $\pm$-boundary condition and coupling constant~$J$.
Let $\mstar=\mstar(J)$ denote the spontaneous magnetization. Then there exist
functions $\mathscr{M}_{\pm,J}\colon[-\mstar,\mstar]\to[0,\infty)$ such that
\begin{equation}
\label{surfLDP}
\lim_{\epsilon\downarrow0}\lim_{L\to\infty}
\frac1{L^{d-1}}\log \BbbP_L^{\pm,J}\bigl(|M_L-mL^d|\le \epsilon L^d\bigr)
=-\mathscr{M}_{\pm,J}(m)
\end{equation}
holds for each $m\in[-\mstar,\mstar]$. Moreover, there is a
constant~$w_1\in(0,\infty)$ such that
\begin{equation}
\label{Meq}
\mathscr{M}_{\pm,J}(m)=\Bigl(\frac{\mstar\mp m}{2\mstar}\Bigr)^{\frac{d-1}d}w_1
\end{equation}
is true for all~$m\in[-\mstar,\mstar]$.  }
\bigskip

The first part of Assumption~A---the surface-order large-deviation principle
\eqref{surfLDP}---has rigorously been verified for square boxes (and
magnetizations near~$\pm\mstar$) in $d=2$~\cite{DKS,Bob+Tim} and in $d \geq
3$~\cite{Bodineau,Cerf-Pisztora}.  The extension to
Wulff-shape domains for all~$m\in[-\mstar,\mstar]$ requires only minor
modifications in $d=2$~\cite{SS1}.  For $d \geq 3$ Wulff-shape
domains should be analogously controllable but explicit details have not
appeared. The fact (proved in \cite{SS1} for $d=2$) that the rate function is
given by \eqref{Meq} for
\emph{all} magnetizations in~$[-\mstar,\mstar]$ is specific to the Wulff-shape
domains; for other domains one expects the formula to be true only
when~$|\mstar\mp m|$ is small enough to ensure that the appropriately-sized
Wulff-shape droplet fits inside the enclosing volume.  Thus, Assumption~A is a proven fact
for $d=2$, and it is imminently provable for $d \geq 3$.

The underlying reason why \eqref{surfLDP} holds is the existence of
multiple states. Indeed, to achieve the magnetization
$m\in(-\mstar,\mstar)$ one
does not have to alter the local distribution of the spin configurations (which
is what has to be done for $m\not\in[-\mstar,\mstar]$); it suffices to create a
\emph{droplet} of one phase inside the other. The cost is just the surface free
energy of the droplet; the best possible droplet is obtained by optimizing the
Wulff functional \eqref{1.7}. This is the content of \eqref{Meq}.
However, the
droplet is confined to a finite set and, once it becomes sufficiently
large, the shape of the enclosing volume becomes relevant.  In generic volumes
the presence of this additional constraint in the
variational problem actually makes the resulting cost \emph{larger}
than~\eqref{Meq}---which represents the cost of an unconstrained
droplet. But, in
Wulff-shape volumes,
\eqref{Meq} holds regardless of the droplet size as long as $|m|\le\mstar$. An
explicit formula for~$\mathscr{M}_{\pm,J}(m)$ for square volumes has been
obtained in $d=2$~\cite{SS2}; the situation in $d\ge3$ has been addressed
in~\cite{GS1,GS2}.

\smallskip
On the basis of the above assumptions, we are ready to
state our first main result concerning the measure~$P_{W_L}^{\pm,c,h}$ with
$c\sim\xi/L$ and $h\sim b/L$.  Using~$\theta$ to denote the fraction of salt on
the plus spins, we begin by introducing the relevant entropy function
\begin{equation}
\label{upsi}
\Upsilon(m,\theta)=-\theta\log\frac{2\theta}{1+m}-(1-\theta)
\log\frac{2(1-\theta)}{1-m}.
\end{equation}
We remark that if we write a full expression for the bulk
entropy,~$\Xi(m,\theta;c)$, see formula \eqref{2.3}, at fixed~$m$, $c$
and~$\theta$, then, modulo some irrelevant terms, the quantity
$\Upsilon(m,\theta)$ is given by $(\partial/\partial c) \Xi(m,\theta;c)$ at
$c=0$. Thus, when we scale~$c\sim\xi/L$, the quantity $\xi\Upsilon(m,\theta)$
represents the relevant (surface order) entropy of salt with~$m$ and~$\theta$
fixed. The following is an analogue of Theorem~\ref{I.thm1} from Part~I for the
case at hand:

\begin{theorem}
\label{thm3} Let $d\ge2$ and let~$J>\Jc(d)$ and~$\kappa>0$ be fixed. Let
$\mstar=\mstar(J)$ denote the spontaneous magnetization of the Ising model.
Suppose that \eqref{surfLDP} in Assumption~A holds and let $(c_L)$
and $(h_L)$ be
two sequences such that $c_L\ge0$ for all~$L$ and that the limits
\begin{equation}
\label{xiblim}
\xi=\lim_{L\to\infty}Lc_L
\quad\text{and}\quad b=\lim_{L\to\infty}Lh_L
\end{equation}
exist and are finite.  Then for all~$m\in[-\mstar,\mstar]$,
\begin{multline}
\label{2.6aa}
\qquad
\lim_{\epsilon\downarrow0}\lim_{L\to\infty}
\frac1{L^{d-1}}\log P_{W_L}^{\pm,c_L,h_L}\bigl(|M_L-mL^d|\le \epsilon L^d\bigr)
\\=-Q_{b,\xi}^\pm(m)+\inf_{|m'|\le\mstar}Q_{b,\xi}^\pm(m'),
\qquad
\end{multline} 
where $Q_{b,\xi}^\pm(m)=\inf_{\theta\in[0,1]}\scrQ_{b,\xi}^\pm(m,\theta)$
with
\begin{equation}
\label{2.7}
\scrQ_{b,\xi}^\pm(m,\theta) =-bm-\xi\kappa\theta-\xi\Upsilon(m,\theta)
+\mathscr{M}_{\pm,J}(m),
\end{equation}
\end{theorem}

Various calculations in the future  will require a somewhat more explicit expression for the rate function~$m\mapsto Q_{b,\xi}^\pm(m)$ on the right-hand side of \eqref{2.6aa}. To derive such an expression, we first note that the minimizer of~$\theta\mapsto\scrQ_{b,\xi}^\pm(m,\theta)$ is uniquely determined by the equation
\begin{equation}
\frac\theta{1-\theta}=\frac{1+m}{1-m}\,e^\kappa.
\end{equation}
Plugging this into~$\scrQ_{b,\xi}^\pm(m,\theta)$ tells us that
\begin{equation} 
Q_{b,\xi}^\pm(m)=-bm-\xi g(m)+\scrM_{\pm,J}(m),
\end{equation}
where
\begin{equation} 
\label{g}
g(m) 
= \log \left( \frac{1-m}{2} + e^\kappa \frac{1+m}{2} \right).
\end{equation}
Clearly,~$g$ is strictly concave for any~$\kappa>0$.

\subsection{Macroscopic phase separation---``liquid'' boundary conditions}
\label{sec2.3}\noindent 
While Theorem~\ref{I.thm1} of~Part~I and Theorem~\ref{thm3} above may appear
formally similar, the solutions of the associated variational
problems are rather different. Indeed, unlike the ``bulk'' rate function $G_{h,c}(m)$ of
Part~I, the functions $Q_{b,\xi}^\pm(m)$ are not generically strictly convex which in turns
leads to a possibility of having more than one minimizing~$m$.  We consider first the case
of plus (that is, liquid water) boundary conditions.

Let $d\ge2$ and let~$J>\Jc(d)$ and~$\kappa>0$ be fixed. To make our formulas manageable, for any function $\phi\colon[-\mstar,\mstar]\to\R$ let us use the abbreviation
\begin{equation}
\label{Dphi}
D^\star_\phi = \frac{\phi(\mstar) -\phi(-\mstar)}{2\mstar}
\end{equation}
for the slope of~$\phi$ between~$-\mstar$ and~$\mstar$.
Further, let us introduce the quantity
\begin{equation}
     \xit = \frac{w_1}{2\mstar d} \bigl( g'(-\mstar) - D^\star_g \bigr)^{-1}
\end{equation}
and the piecewise linear function~$b_2\colon[0,\infty)\to\R$ which is defined by
\begin{equation}
\label{2.13}
     b_2(\xi) = \begin{cases}
     -\frac{w_1}{2\mstar} - \xi D^\star_g, \quad &\xi < \xit 
     \\*[2mm]
     -\frac{d-1}{d}\frac{w_1}{2\mstar} - \xi g'(-\mstar), \quad &\xi \geq \xit.
     \end{cases}
\end{equation}
Our next result is as follows:

\begin{theorem}
\label{thm4} 
Let $d\ge2$ and let~$J>\Jc(d)$ and~$\kappa>0$ be fixed. Let the objects
$Q_{b,\xi}^+$, $\xit$ and~$b_2$ be as defined above.
Then there exists
a (strictly) decreasing and continuous function
$b_1\colon[0,\infty)\to\R$ with the following properties:
\settowidth{\leftmargini}{(1111)}
\begin{enumerate}
\item[(1)]
$b_1(\xi)\ge b_2(\xi)$ for all~$\xi\ge0$, and
$b_1(\xi)=b_2(\xi)$ iff $\xi\le\xit$.
\item[(2)] $b_1'$ is
continuous on $[0,\infty)$, $b_1'(\xi) \to -g'(\mstar)$ as $\xi \to \infty$ and
$b_1$ is strictly convex on $[\xit,\infty)$.
\item[(3)] For~$b\ne b_1(\xi),b_2(\xi)$, the function~$m\mapsto Q_{b,\xi}^+(m)$
is minimized by a single number $m=m_+(b,\xi)\in[-\mstar,\mstar]$
which satisfies
\begin{equation}
\label{2.4a} 
m_+(b,\xi)
\begin{cases} =\mstar,\qquad&\text{if }\,b>b_1(\xi),
\\
\in(-\mstar,\mstar),\qquad&\text{if }\,b_2(\xi)<b<b_1(\xi),
\\ =-\mstar,\qquad&\text{if }\,b<b_2(\xi).
\end{cases}
\end{equation}
\item[(4)] 
The function~$b\mapsto m_+(b,\xi)$ is strictly increasing
for $b \in [b_2(\xi),b_1(\xi)]$, is continuous on the portion of the
line~$b=b_2(\xi)$ for which~$\xi > \xit$ and has a jump discontinuity along
the line defined by~$b=b_1(\xi)$. The only minimizers at~$b=b_1(\xi)$ and~$b=b_2(\xi)$
are the corresponding limits of~$b\mapsto m_+(b,\xi)$.
\end{enumerate}
\end{theorem}

The previous statement essentially characterizes the phase diagram
for the cases
described in \eqref{xiblim}. Focusing on the plus boundary condition
we have the
following facts: For reduced concentrations~$\xi$ exceeding the critical
value~$\xit$, there exists a range of reduced magnetic fields~$b$ where a
non-trivial droplet appears in the system. This range is enclosed by two curves
which are the graphs of functions~$b_1$ and~$b_2$ above. For~$b$ decreasing
to~$b_1(\xi)$, the system is in the pure plus---i.e., liquid---phase but,
interestingly, at~$b_1$
a macroscopic droplet---an ice crystal---suddenly appears in the system. As~$b$
further decreases
the ice crystal keeps growing to subsume the entire system when~$b=b_2(\xi)$.
For~$\xi\le\xit$ no phase separation occurs; the transition at~$b=b_1(\xi)=b_2(\xi)$ is directly from~$m=\mstar$ to~$m=-\mstar$.

It is noted that the situation for~$\xi$ near
zero corresponds to the Ising model with negative external field proportional
to~$1/L$. In two-dimensional setting, the latter problem has been studied
in~\cite{SS1}.   As already mentioned, the generalizations to rectangular boxes
will require a non-trivial amount of extra work. For the unadorned
Ising model
(i.e.,~$c=0$) this has been carried out in great detail in~\cite{SS2} for~$d=2$
(see also~\cite{Kotecky-Medved}) and in less detail in general
dimensions~\cite{GS1,GS2}.

\smallskip
It is reassuring to observe that the above results mesh favorably with the corresponding asymptotic of Part~I. For finite concentrations and external fields, there are two curves,~$c\mapsto h_+(c)$ and~$c\mapsto h_-(c)$, which mark the boundaries of the phase separation region against the liquid and ice regions, respectively. The curve~$c\mapsto h_+(c)$ is given by the equation
\begin{equation}
     h_+(c) = \frac{1}{2} \log \frac{1-q_+}{1-q_-},
\end{equation}
where $(q_+,q_-)$ is the (unique) solution of
\begin{equation}
    \frac{q_+}{1-q_+} = e^\kappa \frac{q_-}{1-q_-}, \qquad
     q_+ \frac{1+\mstar}{2} + q_- \frac{1-\mstar}{2} = c.
\end{equation}
The curve $c\mapsto h_-(c)$ is defined by the same equations with the roles of $\mstar$ and $-\mstar$ interchanged. Since $h_\pm(0) = 0$, these can be linearized around the point~$(0,0)$. Specifically, plugging~$b/L$ for~$h$ and~$\xi/L$ for~$c$ into~$h=h_\pm(c)$ and letting~$L\to\infty$ yields the linearized versions
\begin{equation}
b_\pm= h_\pm'(0)\xi
\end{equation}
of~$h_+$ and~$h_-$. It is easy to check that~$h_\pm'(0) = -g'(\pm \mstar)$ and so, in the limit~$\xi\to\infty$, the linear function~$b_+$ has the same slope as~$b_1$ while~$b_-$ has the same slope as~$b_2$ above. Theorem~\ref{thm4} gives a detailed description of how these linearized curves ought to be continued into (infinitesimal) neighborhoods of size~$1/L$ around~$(0,0)$.

\subsection{Macroscopic phase separation---``ice'' boundary conditions}
\label{sec2.4}\noindent 
Next we consider minus (ice) boundary conditions, where the requisite liquid
water, phase separation and ice regions will be defined using the functions $\tb_1 \geq
\tb_2$.  As for the plus boundary conditions, there is a value
$\txit > 0$ where the phase separation region begins, but now we have a new
phenomenon:  For some (but not all) choices of
$J$ and~$\kappa$, there exists a nonempty interval $(\txit,\txiu)$ of~$\xi$ for which two distinct droplet transitions occur. Specifically, as~$b$
increases, the volume fraction occupied by the droplet first jumps
discontinuously
at $\tb_2(\xi)$ from zero to a strictly positive value, then increases but stays
bounded away from one,  and then, at~$b=\tb_1(\xi)$, jumps discontinuously to one; i.e., the ice surrounding the droplet suddenly melts.

\smallskip
For each~$J>\Jc(d)$ and each~$\kappa$, consider the auxiliary quantities
\begin{equation}
\label{thexis}
     \xi_1 = \frac{w_1}{2\mstar d} \bigl( D^\star_g - g'(\mstar) \bigr)^{-1}
     \quad\text{and}\quad
     \xi_2 = -\frac{(d-1)w_1}{(2\mstar d)^2 g''(\mstar)}.
\end{equation}
(Note that, due to the concavity property of~$g$, both~$\xi_1$ and~$\xi_2$ are finite and positive.) 
The following is a precise statement of the above:

\begin{theorem}
\label{thm4b} 
Let $d\ge2$ and let~$J>\Jc(d)$ and~$\kappa>0$ be fixed. Then there exist two
(strictly) decreasing and continuous functions $\tb_1,\tb_2\colon[0,\infty)\to\R$ and numbers $\txit,\txiu\in (0,\infty)$ with $\txit \leq \txiu$ such that the following properties hold:
\settowidth{\leftmargini}{(1111a)}
\begin{enumerate}
\item[(1)]
$\tb_1(\xi)\ge \tb_2(\xi)$ for all~$\xi\ge0$, and
$\tb_1(\xi)=\tb_2(\xi)$ iff $\xi\le\txit$.
\item[(2)]
$\tb_2$ is strictly concave on $[\txit,\infty)$,
$\tb_2'(\xi) \to -g'(-\mstar)$ as $\xi \to \infty$,
$\tb_1$ is strictly convex on $(\txit,\txiu)$ and, outside this interval,
\begin{equation}
\label{2.19}
     \tb_1(\xi) = \begin{cases}
     \frac{w_1}{2\mstar} - \xi D^\star_g, \quad &\xi \leq \txit,
     \\*[2mm]
     \frac{d-1}{d}\frac{w_1}{2\mstar} - \xi g'(\mstar), \quad &\xi 
\geq \txiu. \\
     \end{cases}
\end{equation}
\item[(3a)] 
If $\xi_1\geq \xi_2$, 
then $\txit = \txiu = \xi_1$ and $\tb_2'$ is
continuous on $[0,\infty)$.
\item[(3b)] 
If $\xi_1<\xi_2$ then $\txit < \xi_1 < \txiu = \xi_2$ and neither~$b_1'$ nor~$b_2'$ is continuous at~$\txit$. 
Moreover, there exists $m_0\in (-\mstar,\mstar)$ such that, as~$\xi\downarrow \txit$,
\begin{equation} 
\label{derivs}
   b_1'(\xi) \to -\frac{g(\mstar) - g(m_0)}{\mstar - m_0} 
   \quad\text{and}\quad
   b_2'(\xi) \to -\frac{g(m_0) - g(-\mstar)}{m_0 + \mstar}.
\end{equation}
\item[(4)] 
For~$b\ne \tb_1(\xi),\tb_2(\xi)$, the function~$m\mapsto
Q_{b,\xi}^-(m)$ is minimized by a single number
$m=m_-(b,\xi)\in[-\mstar,\mstar]$
which satisfies
\begin{equation}
\label{2.4b} 
m_-(b,\xi)\begin{cases} =\mstar,\qquad&\text{if }\,b>\tb_1(\xi),
\\
\in(-\mstar,\mstar),\qquad&\text{if }\,\tb_2(\xi)<b<\tb_1(\xi),
\\ =-\mstar,\qquad&\text{if }\,b<\tb_2(\xi).
\end{cases}
\end{equation}
\item[(5)] 
The function~$b\mapsto m_-(b,\xi)$ is strictly increasing in~$b$
for $b \in [\tb_2(\xi),\tb_1(\xi)]$, is continuous on the portion of the
line~$b=\tb_1(\xi)$ for which~$\xi \geq \txiu$ and has jump discontinuities
both along the line defined by~$b=\tb_2(\xi)$ and along the portion of the
line $b = \tb_1(\xi)$ for which $\txit < \xi < \txiu$. There are two minimizers
at the points where~$b\mapsto m_-(b,\xi)$ is discontinuous with the exception of~$(b,\xi)=(\tb_1(\txit),\txit)=(\tb_2(\txit),\txit)$ when~$\txit<\txiu$, where there are three minimizers; namely,~$\pm\mstar$ and~$m_0$ from part~(3b). 
\end{enumerate}
\end{theorem}

As a simple consequence of the definitions, it is seen that the question of whether or not~$\xi_1\ge\xi_2$ is equivalent to the question whether or not
\begin{equation} 
\label{gapcond}
     g(\mstar) - 2\mstar g'(\mstar) + \frac{d}{d-1}(2\mstar)^2 g''(\mstar)
     \leq g(-\mstar).
\end{equation}
We claim that \eqref{gapcond} will hold, or fail, depending on the values of the various parameters of the model. Indeed, writing~$\epsilon=\tanh(\kappa/2)$ we get
\begin{equation}
g(m)=\log(1+\epsilon m)+\text{const.}
\end{equation}
Regarding the quantity~$\epsilon m$ as a ``small parameter,'' we easily verify that the desired inequality holds to the lowest non-vanishing order. Thus, if~$\mstar$ is small enough, then \eqref{gapcond} holds for all~$\kappa$, while it is satisfied for all~$\mstar$ whenever~$\kappa$ is small enough. On the other hand, as~$\kappa$ tends to infinity, $g(\mstar)-g(-\mstar)$ tends to~$\log\frac{1+\mstar}{1-\mstar}$, while the various relevant derivatives of~$g$ are bounded independently of~$\mstar$. Thus, as~$\mstar\to1$, which happens when~$J\to\infty$, the condition \eqref{gapcond} is \emph{violated} for~$\kappa$ large enough. Evidently,
the gap $\txiu-\txit$ is strictly positive for some choices of~$J$ and~$\kappa$, and vanishes for others.

Since $\tb_1(0) > 0$, for~$\xi$ sufficiently small the ice region includes points with~$b>0$ .  Let us also show that the phase separation region can rise above~$b=0$; as indicated in the plot on the right of Fig.~\ref{fig2}. Clearly, it suffices to consider~$b=0$ and
establish that for some $J$, $\kappa$ and~$\xi$, the absolute minimum of $m\mapsto Q_{0,\xi}^-(m)$ does not occur at $\pm \mstar$. This will certainly hold if
\begin{equation}
     (Q_{0,\xi}^-)'(\mstar) > 0 
     \quad\text{and}\quad 
     Q_{0,\xi}^-(-\mstar)>Q_{0,\xi}^-(\mstar),
\end{equation}
or, equivalently, if
\begin{equation} \label{b0conds}
     \frac{d-1}{d} \frac{w_1}{2\mstar} > \xi g'(\mstar)
     \quad\text{and}\quad
     \xi\bigl( g(\mstar) - g(-\mstar) \bigr) > w_1
\end{equation}
are both true. Some simple algebra shows that the last inequalities hold for \emph{some}~$\xi$ once
\begin{equation}
\frac{d-1}d\bigl( g(\mstar) - g(-\mstar) \bigr)>2\mstar g'(\mstar).
\end{equation}
But, as we argued a moment ago, the difference $g(\mstar) - g(-\mstar)$ can be made arbitrary large by taking $\kappa\gg1$ and~$\mstar$ sufficiently close to one, while~$g'(\mstar)$ is bounded in these limits. So, indeed, the phase separation region pokes above the~$b=0$ axis once~$\kappa\gg1$ and~$J\gg1$. 

Comparing to the linear asymptotic of the phase diagram from Part~I, we see that in the
finite-volume system with minus (ice) boundary condition, the lines
bounding the phase separation region are shifted upward and again are pinched together.  
In this case it is the line $b = \tb_1(\xi)$ that is parallel to its counterpart $b
= h_+'(0)\xi$ for $\xi > \txiu$, while $b = \tb_2(\xi)$ has the same asymptotic slope (in the limit $\xi \to \infty$) as the function $b = h_-'(0)\xi$.

\subsection{Properties of the spin marginal}
\noindent
On the basis of Theorems~\ref{thm3}--\ref{thm5}, we can
now provide a routine characterization of the typical configurations in
measure~$P_{W_L}^{\pm,c_L,h_L}$. The following is an analogue of Theorem~2.2 of
Part~I for the cases at~hand:
\begin{theorem}
\label{thm5} Let $d\ge2$ and let~$J>\Jc(d)$ and~$\kappa>0$ be fixed.
Suppose that
Assumption~A holds and let $(c_L)$ and $(h_L)$ be two sequences such that
$c_L\ge0$ for all~$L$ and that the limits~$\xi$ and~$b$ in \eqref{xiblim} exist
and are finite. Let us define two sequences of Borel probability
measures~$\rho_L^\pm$ on~$[-\mstar,\mstar]$ by putting
\begin{equation}
\rho_L^\pm\bigl([-1,m]\bigr)=P_{W_L}^{\pm,c_L,h_L}(M_L\le mL^d),
\qquad m\in[-1,1].
\end{equation}
Then the spin marginal of the measure~$P_{W_L}^{\pm,c_L,h_L}$ can again be
written as a convex combination of the Ising measures with fixed magnetization;
i.e., for any set~$\AA$ of configurations $(\sigma_x)_{x\in\Lambda_L}$,
\begin{equation}
\label{2.6b}
P_{W_L}^{\pm,c_L,h_L}\bigl(\AA\times\{0,1\}^{W_L}\bigr)=\int\!\rho_L^\pm(\textd
m)\,
\BbbP_{W_L}^{\pm,J}\bigl(\AA\big|M_L=\lfloor mL^d\rfloor\bigr).
\end{equation}
Moreover, any (weak) subsequential limit~$\rho^\pm$ of
measures~$\rho_L^\pm$ is concentrated on the minimizers of $m\mapsto
Q_{b,\xi}^\pm(m)$. In particular, for~$b\ne b_1(\xi),b_2(\xi)$ the limit
$\rho^+=\lim_{L\to\infty}\rho_L^+$ exists and is simply the Dirac mass
at~$m_+(b,\xi)$---the quantity from Theorem~\ref{thm4}---and similarly for
$\rho^-=\lim_{L\to\infty}\rho_L^-$ and $b\ne\tb_1(\xi),\tb_2(\xi)$.
\end{theorem}

On the basis of Theorems~\ref{thm3}--\ref{thm5}, we can draw the
following conclusions: For $d$-dimensional systems of scale~$L$ with the total
amount of salt proportional to $L^{d-1}$ (i.e., the system boundary), phase
separation occurs \emph{dramatically} in the sense that all of a sudden a
non-trivial fraction of the system melts/freezes (depending on the boundary
condition). In hindsight, this is perhaps not so difficult to
understand. While
a perturbation of size $L^{d-1}$ cannot influence the bulk properties of the
system with a single phase, here the underlying system is at phase coexistence.
Thus the cost of a droplet is only of order~$L^{d-1}$, so it is
not unreasonable that a comparable amount of salt will cause dramatic effects.  

It is worth underscoring that
the jump in the size of the macroscopic droplet at~$b=b_1$ or~$b=\tb_2$ decreases with increasing~$\xi$.
Indeed, in the extreme limit, when the concentration is finite
(nonzero) we know
that no macroscopic droplet is present at the transition. But, presumably, by
analogy with the results of \cite{bigBCK} (see also~\cite{BCK,Neuhaus-Hager}),
there will be a \emph{mesoscopic} droplet---of a particular scaling---appearing at the
transition point. This suggests that a host of intermediate
mesoscopic scales may
be exhibited depending on how~$c_L$ and~$h_L$ tend to zero with the ratio~$h_L/c_L$
approximately fixed. These intermediate behaviors are currently being
investigated.

\section{Proofs of main results}
\label{sec3}\noindent 
The goal of this section is to prove the results stated in
Section~\ref{sec2}. We begin by stating a generalized large deviation principle
for both magnetization and the fraction of salt on the plus spins from which
Theorem~\ref{thm3} follows as an easy corollary. Theorem~\ref{thm4} is proved in Section~\ref{sec3.2}; Theorems~\ref{thm4b} and~\ref{thm5} are proved in Section~\ref{sec3.3}.

\subsection{A generalized large-deviation principle} 
\noindent
We will proceed
similarly as in the proof of Theorem~\ref{I.thm3.5} from~Part~I. Let
$\Lambda\subset\Z^d$ be a
finite set and let us reintroduce the quantity
\begin{equation}
Q_\Lambda=\sum_{x\in\Lambda}\sS_x\frac{1+\sigma_x}2,
\end{equation}
which gives the total amount of salt on the plus spins
in~$\Lambda$.
Recall that~$\E_\Lambda^{\pm,J}$ denotes the expectation with respect to the
(usual) Ising measure with coupling constant~$J$ and plus/minus boundary
conditions.
First we generalize a couple of statements from~Part~I:

\begin{lemma}
\label{lemma3.2} Let~$\Lambda\subset\Z^d$ be a finite set. Then for any fixed
spin configuration~$\bar\sigma=(\bar\sigma_x)\in\{-1,1\}^\Lambda$, all salt
configurations $(\sS_x)\in\{0,1\}^\Lambda$ with the same~$N_\Lambda$
and~$Q_\Lambda$ have the same probability in the conditional
measure~$P_\Lambda^{\pm,c,h}(\ccdot|\sigma=\bar\sigma)$. Moreover, for
any~$\barsS=(\barsS_x)\in\{0,1\}^\Lambda$ with $N_\Lambda=\lfloor
c|\Lambda|\rfloor$ and for any~$m\in[-1,1]$,
\begin{multline}
\label{3.3}
\quad
P_\Lambda^{\pm,c,h}\bigl(\,\barsS\,\text{\rm~occurs},\,M_\Lambda=\lfloor m|\Lambda|\rfloor\bigr) 
\\=
\frac1{Z_\Lambda}\E_\Lambda^{\pm,J}\bigl(e^{\kappa Q_\Lambda(\sigma,\barsS)+hM_\Lambda(\sigma)}
\1_{\{M_\Lambda(\sigma)=\lfloor m|\Lambda|\rfloor\}}\bigr),
\quad
\end{multline}
where the normalization constant is given by
\begin{equation}
\label{3.4}
Z_\Lambda=\sum_{\sS'\in\{0,1\}^\Lambda}\1_{\{N_\Lambda(\sS')=\lfloor
c|\Lambda|\rfloor\}}\,
\E_\Lambda^{\pm,J}\bigl(e^{\kappa
Q_\Lambda(\sigma,\sS')+hM_\Lambda(\sigma)}\bigr).
\end{equation}
\end{lemma}

\begin{proofsect}{Proof} 
This is identical to Lemma~\ref{I.lemma3.2}
from~Part~I.
\end{proofsect}

Next we will sharpen the estimate from~Part~I concerning the total entropy
carried by the salt. Similarly to the object~$\AA_L^{\theta,c}(\sigma)$
from~Part~I, for each spin configuration~$\sigma=(\sigma_x)\in\{-1,1\}^\Lambda$
and numbers~$\theta,c\in[0,1]$, we introduce the set
\begin{equation}
\AA_\Lambda^{\theta,c}(\sigma)=\bigl\{(\sS_x)\in\{0,1\}^\Lambda\colon
N_L=\lfloor
c|\Lambda|\rfloor,\,Q_L=\lfloor\theta c|\Lambda|\rfloor\bigr\}.
\end{equation}
Clearly, the \emph{size} of~$\AA_\Lambda^{\theta,c}(\sigma)$ is
the same for all~$\sigma$ with a given value of the magnetization; we will thus
let~$A_\Lambda^{\theta,c}(m)$ denote the common value
of~$|\AA_\Lambda^{\theta,c}(\sigma)|$ for those~$\sigma$
with~$M_\Lambda(\sigma)=\lfloor m|\Lambda|\rfloor$. Let $\scrS(p)=p\log
p+(1-p)\log(1-p)$ and let us recall the definition of the entropy function
\begin{equation}
\label{2.3}
\Xi(m,\theta;c)= -\frac{1+m}2\scrS\Bigl(\frac{2\theta c}{1+m}\Bigr)
-\frac{1-m}2\scrS\Bigl(\frac{2(1-\theta)c}{1-m}\Bigr);
\end{equation}
 cf~formula \eqref{I.2.3} from Part~I. Then we have:

\begin{lemma}
\label{lemma3.3} For each~$\eta>0$ there exist constants~$C_1<\infty$
and $L_0<\infty$ such that for all finite~$\Lambda\subset\Z^d$
with~$|\Lambda|\ge
L_0^d$, all $\theta,c\in[0,1]$ and all~$m$ with~$|m|\le1-\eta$ satisfying
\begin{equation}
\label{3.5}
\frac{2\theta
c}{1+m}\le1-\eta\quad\text{and}\quad\frac{2(1-\theta)c}{1-m}\le1-\eta
\end{equation}
we have
\begin{equation}
\label{3.6}
\biggl|\frac{\log A_\Lambda^{\theta,c}(m)}{|\Lambda|}-\Xi(m,\theta;c)\biggr|\le
C_1\frac{\log|\Lambda|}{|\Lambda|}.
\end{equation}
\end{lemma}

\begin{proofsect}{Proof} 
The same calculations that were used in the proof of
Lemma~\ref{I.lemma3.3} from~Part~I give us
\begin{equation}
\label{3.8a}
A_\Lambda^{\theta,c}(m)=\binom{\frac12(|\Lambda|+M_\Lambda)}
{Q_\Lambda}\binom{\frac12(|\Lambda|-M_\Lambda)}{N_\Lambda-Q_\Lambda}
\end{equation}
with the substitutions~$M_\Lambda=\lfloor m|\Lambda|\rfloor$ and~$Q_\Lambda=\lfloor\theta c|\Lambda|\rfloor$.
By \eqref{3.5} and~$|m|\le1-\eta$, both combinatorial
numbers are
well defined once~$|\Lambda|$ is sufficiently large (this defines~$L_0$). Thus,
we can invoke the Stirling approximation and, eventually, we see that the
right-hand side of \eqref{3.8a}
equals~$\exp\{|\Lambda|\Xi(m,\theta;c)\}$ times
factors which grow or decay at most like a power of~$|\Lambda|$.
Taking logs and
dividing by~$|\Lambda|$, this yields \eqref{3.6}.
\end{proofsect}

Our final preliminary lemma is concerned with the magnetizations outside 
$[-\mstar,\mstar]$ which are (formally) not covered by Assumption~A.
Recall the sequence of Wulff shapes~$W_L$ defined at the end of
Section~\ref{sec1.2}. Note that~$W_L$ contains, to within boundary
corrections,~$L^d$ sites.

\begin{lemma}
\label{lemma3.4} 
Suppose that~$J>\Jc$ and let~$c_L$ and~$h_L$ be such
that~$Lc_L$
and~$Lh_L$ have finite limits as~$L\to\infty$. For each~$\epsilon>0$, we have
\begin{equation}
\label{3.9c}
\lim_{L\to\infty}\frac1{L^{d-1}}\log
P_{W_L}^{\pm,c_L,h_L}\bigl(|M_{W_L}|\ge(\mstar+\epsilon)L^d\bigr)=-\infty.
\end{equation}
\end{lemma}

\begin{proofsect}{Proof} 
This is a simple consequence of the fact that, in the unadorned Ising magnet, the probability in \eqref{3.9c} is exponentially small in \emph{volume}---cf Theorem~\ref{I.thm3.1}---and that with~$Lh_L$ and~$Lc_L$ bounded, there will be at most a surface-order correction. A formal proof proceeds as follows: We write
\begin{equation}
\label{3.7b} 
P_{W_L}^{\pm,c_L,h_L}\bigl(Q_L=\lfloor\theta
c_LL^d\rfloor,\,M_L=\lfloor mL^d\rfloor\bigr)=\frac{\widetilde
K_L(m,\theta)}{Y_L},
\end{equation}
where
\begin{equation}
\label{3.8}
\widetilde K_L(m,\theta)=A_{W_L}^{\theta,c_L}(m)\,e^{h_L\lfloor
mL^d\rfloor+\kappa\lfloor\theta c_LL^d\rfloor}\,
\BbbP_{W_L}^{\pm,J}\bigl(M_L=\lfloor mL^d\rfloor\bigr)
\end{equation}
and where~$Y_L$ is the sum of~$\widetilde K_L(m',\theta')$ over all relevant values of~$m'$ and~$\theta'$.
Under the assumption that both~$h_L$ and~$c_L$ behave
like~$O(L^{-1})$, the prefactors of the Ising probability can be bounded
between~$e^{-CL^{d-1}}$ and $e^{CL^{d-1}}$, for some~$C<\infty$, uniformly
in~$\theta$ and~$m$. This yields
\begin{equation}
\label{3.12b}
P_{W_L}^{\pm,c_L,h_L}\bigl(|M_{W_L}|\ge(\mstar+\epsilon)L^d\bigr)\le
e^{CL^{d-1}}\,\frac1{Y_L}
\BbbP_{W_L}^{\pm,J}\bigl(|M_{W_L}|\ge(\mstar+\epsilon)L^d\bigr).
\end{equation}
The same argument shows us that~$Y_L$ can be bounded below by~$e^{-CL^{d-1}}$
times the probability that~$M_{W_L}$ is near zero in the Ising
measure~$\BbbP_{W_L}^{\pm,J}$. 
In light of~$J>\Jc$, Assumption~A then gives
\begin{equation}
\liminf_{L\to\infty}\frac1{L^{d-1}}\log Y_L>-\infty.
\end{equation}
On the other hand, by Theorem~\ref{I.thm3.1} (and the remark that
follows it) we have that
\begin{equation}
\lim_{L\to\infty}\frac1{L^{d-1}}\log
\BbbP_{W_L}^{\pm,J}\bigl(|M_{W_L}|\ge(\mstar+\epsilon)L^d\bigr)=-\infty.
\end{equation}
Plugging this into \eqref{3.12b}, the desired claim follows.
\end{proofsect}

We will use the above lemmas to state and prove a generalization of
Theorem~\ref{thm3}.
\begin{theorem}
\label{thm3.5} Let~$d\ge2$ and let~$J>\Jc(d)$ and~$\kappa\ge0$ be fixed.
Let~$c_L\in[0,1]$ and~$h_L\in\R$ be two sequences such that the limits~$\xi$
and~$b$ in \eqref{xiblim} exist and are finite. For each~$m\in[-\mstar,\mstar]$
and $\theta\in(-1,1)$,
let~$\widetilde\BB_{L,\epsilon}=\widetilde\BB_{L,\epsilon}(m,c_L,\theta)$
be the
set of all $(\sigma,\sS)\in\{-1,1\}^{W_L}\times\{0,1\}^{W_L}$ for
which the bounds
\begin{equation}
|M_{W_L}-mL^d|\le\epsilon L^d
\quad\text{and}\quad |Q_{W_L}-\theta c_LL^d|\le\epsilon L^{d-1}
\end{equation}
hold. Then
\begin{equation}
\label{3.10}
\lim_{\epsilon\downarrow0}\lim_{L\to\infty}\frac{\log
P_{W_L}^{\pm,c_L,h_L}(\widetilde\BB_{L,\epsilon})}{L^{d-1}}=
-\scrQ_{b,\xi}(m,\theta)+\inf_{\begin{subarray}{c}
|m'|\le\mstar\\\theta'\in[0,1]
\end{subarray}}
\scrQ_{b,\xi}(m',\theta'),
\end{equation}
where~$\scrQ_{b,\xi}(m,\theta)$ is as in \eqref{2.7}.
\end{theorem}

\begin{proofsect}{Proof} 
We again begin with the representation
\twoeqref{3.7b}{3.8} for the choices~$h_LL^d\sim bL^{d-1}$ and~$c_LL^d\sim \xi
L^{d-1}$. For~$m\in[-\mstar,\mstar]$ the last probability  in
\eqref{3.8} can be
expressed from Assumption~A and so the only thing to be done is the
extraction of
the exponential rate of~$A_{W_L}^{\theta,c_L}(m)$ to within errors of
order~$o(L^{d-1})$. This will be achieved Lemma~\ref{lemma3.3}, but
before doing
that, let us express the leading order behavior of the
quantity~$\Xi(m,\theta;c_L)$. Noting the expansion~$\scrS(p)=p\log p-p+O(p^2)$
for~$p\downarrow0$ we easily convince ourselves that
\begin{equation}
\begin{aligned}
\Xi(m,\theta;c_L)&=-\theta c_L\Bigl(\log\frac{2\theta c_L}{1+m}-1\Bigr)
\\
&\qquad\quad
-(1-\theta) c_L\Bigl(\log\frac{2(1-\theta)c_L}{1-m}-1\Bigr)+O(c_L^2)
\\&=c_L-c_L\log c_L+c_L\Upsilon(m,\theta)+O(c_L^2),
\end{aligned}
\end{equation}
where~$\Upsilon(m,\theta)$ is as in \eqref{upsi}. (The
quantity~$O(c_L^2)$ is bounded by a constant times~$c_L^2$ uniformly in~$m$
satisfying~$|m|\le1-\eta$ and \eqref{3.5}.) Invoking
Lemma~\ref{lemma3.3} and the
facts that~$|W_L|-L^d=O(L^{d-1})$ and~$Lc_L^2\to0$ as $L\to\infty$ we
now easily
derive that
\begin{equation}
A_{W_L}^{\theta,c_L}(m)=\exp\Bigl\{\,r_L+L^{d-1}\xi\Upsilon(m,\theta)+o(L^{d-1})\Bigr\},
\end{equation}
where~$r_L=-L|W_L|c_L\log(c_L/e)$ is a quantity
independent of~$m$
and~$\theta$.

Putting the above estimates together, we conclude that
\begin{equation}
\widetilde
K_L(m,\theta)=\exp\Bigl\{\,r_L-L^{d-1}\scrQ_{b,\xi}(m,\theta)+o(L^{d-1})\Bigr\}
\end{equation}
where~$o(L^{d-1})$ is small---relative to~$L^{d-1}$---uniformly
in~$m\in[-\mstar,\mstar]$ and~$\theta\in[0,1]$.
It remains to use this expansion to produce the leading order asymptotics
of~$P_{W_L}^{\pm,c_L,h_L}(\widetilde\BB_{L,\epsilon})$. Here we write
the latter
quantity as a ratio,
\begin{equation}
\label{ratio}
P_{W_L}^{\pm,c_L,h_L}(\widetilde\BB_{L,\epsilon})=\frac{\widetilde
K_{L,\epsilon}(m,\theta)}{Y_L},
\end{equation}
where $\widetilde K_{L,\epsilon}(m,\theta)$ is the sum of
$\widetilde K_L(m',\theta')$ over all relevant values
of~$(m',\theta')$ that can
contribute to the event~$\widetilde\BB_{L,\epsilon}$, while, we remind the reader,~$Y_L$ is the sum of~$\widetilde K_L(m',\theta')$ over all relevant~$(m',\theta')$'s
regardless of their~worth. 

It is intuitively clear that the~$r_L$-factors in the
numerator and
denominator cancel out and one is left only with terms of
order~$L^{d-1}$, but to
prove this we will have to invoke a (standard)
compactness argument.  
We first note that for each~$\delta>0$
and each~$(m,\theta)\in[-\mstar,\mstar]\times[0,1]$, there exists
an~$\epsilon>0$
and an~$L_0<\infty$---both possibly depending on~$m$,~$\theta$
and~$\delta$---such that, for~$L\ge L_0$,
\begin{equation}
\label{3.21}
\Bigl|\frac1{L^{d-1}}\log\bigl(\widetilde
K_{L,\epsilon}(m,\theta)e^{-r_L}\bigr)+\scrQ_{b,\xi}(m,\theta)\Bigr|\le\delta.
\end{equation}
(Here we also used that~$\scrQ_{b,\xi}(m,\theta)$ is continuous in
both variables
on~$[-\mstar,\mstar]\times[0,1]$.) 
By compactness of $[-\mstar,\mstar]\times[0,1]$, there exists a finite set
of~$(m_k,\theta_k)$'s such that the above
$\epsilon$-neighboorhoods---for which \eqref{3.21} holds with the
same~$\delta$---cover  the set $[-\mstar,\mstar]\times[0,1]$. In fact we cover
the slightly larger set
\begin{equation}
\RR=[-\mstar-\epsilon',\mstar+\epsilon']\times[0,1],
\end{equation}
where $\epsilon'>0$. By choosing the~$\epsilon$'s sufficiently small, we can
also ensure that for one of the~$k$'s, the quantity~$\scrQ_{b,\xi}(m_k,\theta_k)$ is
within~$\delta$ of its absolute minimum. Since everything is finite, all
estimate are uniform in~$L\ge L_0$ on~$\RR$.  

To estimate~$Y_L$ we will split it 
into two parts,~$Y_{L,1}$ and~$Y_{L,2}$, according to whether the
corresponding~$(m',\theta')$ belongs to~$\RR$ or not. By \eqref{3.21} and the
choice of the above cover of~$\RR$ we have that~$\frac1{L^{d-1}}\log
Y_{L,1}$ is within, say,~$3\delta$ of the minimum
of~$(m,\theta)\mapsto\scrQ_{b,\xi}(m,\theta)$ once~$L$ is
sufficiently large. (Here the additional~$\delta$ is used to control the number of
terms in the cover of~$\RR$.) On
the other hand, Lemma~\ref{lemma3.4} implies that~$Y_{L,2}$ is exponentially
small relative to~$Y_{L,1}$. Hence we get
\begin{equation}
\limsup_{L\to\infty}\,\Bigl|\frac1{L^{d-1}}\log\bigl(Y_Le^{-r_L}\bigr)+
\inf_{\begin{subarray}{c} |m'|\le\mstar\\\theta'\in[0,1]
\end{subarray}}
\scrQ_{b,\xi}(m',\theta')\Bigr|\le3\delta.
\end{equation}
Plugging these into \eqref{ratio} the claim follows by
letting~$\delta\downarrow0$.
\end{proofsect}

\begin{proofsect}{Proof of Theorem~\ref{thm3}} 
This is a simple consequence of
the compactness argument invoked in the last portion of the previous proof.
\end{proofsect}

\subsection{Proof of Theorem~\ref{thm4}}
\label{sec3.2}\noindent 
Here we will prove Theorem~\ref{thm4} which describes the phase diagram for the ``liquid'' boundary condition, see the plot on the left of Fig.~1. 

\begin{proofsect}{Proof of part~(1)}
Our goal is to study the
properties of the function~$m\mapsto Q_{b,\xi}^+(m)$.
Throughout the proof we will keep $J$ fixed (and larger than~$\Jc$) and write $\scrM(\cdot)$ instead of $\scrM_{+,J}(\cdot)$.  For $m\in[-\mstar,\mstar]$, let us define the quantity
\begin{equation}
\label{3.24}
     E_\xi(m) = -\xi g(m) +\scrM(m). 
\end{equation}
Clearly, this is just $Q_{b,\xi}^+(m)$ without the~$b$-dependent part, i.e., $Q_{b,\xi}^+(m)=-bm +E_\xi(m)$. Important for this proof will be the ``zero-tilt'' version of this function,
\begin{equation}
\label{3.25}
     \oE_\xi(m) = E_\xi(m) - E_\xi(-\mstar) - (m+\mstar)D_{E_\xi}^\star,
\end{equation}
where $D_{E_\xi}^\star$ is the ``slope of~$E_\xi$ between~$-\mstar$ and~$\mstar$,'' see \eqref{Dphi}.
Clearly, $E_\xi$ and~$\oE_\xi$ have the same convexity/concavity properties but $\oE_\xi$ always satisfies $\oE_\xi(-\mstar)=\oE_\xi(\mstar) = 0$.  

Geometrically, the minimization of $Q_{b,\xi}^+(m)$ may now be viewed as follows: 
Consider the set of points $\{(m,y)\colon y=E_\xi(m)\}$---namely, the graph of~$E_\xi(m)$---and take the lowest vertical translate of the line $y=bm$ which contacts this set. 
Clearly, the minimum of~$Q_{b,\xi}^+(m)$ is achieved at the value(s) of~$m$ where this contact
occurs. The same of course holds for the graph $y = \oE_\xi(m)$ provided we shift~$b$ by $D_{E_\xi}^\star$. Now the derivative $\oE_\xi'(m)$ is bounded below at~$m=-\mstar$ and above at~$m=\mstar$ (indeed, as~$m\uparrow\mstar$ the derivative diverges to~$-\infty$). It follows that there exist two values, $-\infty < b_1(\xi) \leq b_2(\xi) < \infty$, such that $m=\mstar$ is the unique minimizer for~$b>b_1(\xi)$, $m=-\mstar$ is the unique minimizer for~$b<b_2(\xi)$, and neither~$m=\mstar$ nor~$m=-\mstar$ is a minimizer when~$b_2(\xi)<b<b_1(\xi)$.  

On the basis of the above geometrical considerations, the region where~$b_1$ and~$b_2$ are the same is easily characterized:
\begin{equation} 
\label{b1b2}
    b_1(\xi) = b_2(\xi) \quad \text{if and only if } \quad \oE_\xi(m)
    \geq 0 \quad\forall m\in[-\mstar,\mstar].
\end{equation}
To express this condition in terms of~$\xi$, let us define $T(m) = \scrM''(m)/g''(m)$ and note that $E_\xi''(m) > 0$ if and only if $T(m)>\xi$. Now, for some constant $C = C(J)>0$,
\begin{equation}
     T(m) = C(\mstar - m)^{-\frac{d+1}d}\bigl(m + \cot(\kappa/2)\bigr)^2,
\end{equation}
which implies that~$T$ is strictly increasing on $[-\mstar,\mstar)$ with~$T(m)\to\infty$ as~$m\uparrow\mstar$.  It follows that either~$\oE_\xi$ is concave throughout~$[-\mstar,\mstar]$, or there exists a $T^{-1}(\xi)\in(-\mstar,\mstar)$ such that~$\oE_\xi$ is strictly convex on $[-\mstar,T^{-1}(\xi))$ and strictly concave on $(T^{-1}(\xi),\mstar]$.  Therefore, by \eqref{b1b2}, $b_1(\xi)<b_2(\xi)$ if and only
if $\oE_\xi'(-\mstar) < 0$, which is readily verified to be equivalent to $\xi>\xit$.  This proves part~(1) of the theorem.  
\end{proofsect}

\begin{proofsect}{Proof of parts~(3) and~(4)}
The following properties, valid for~$\xi>\xit$, are readily verified on the basis of the above convexity/concavity picture:
\settowidth{\leftmargini}{(aa)}
\begin{enumerate}
\item[(a)] 
For all $b_2(\xi) < b < b_1(\xi)$, there is a unique minimizer $m_+(b,\xi)$
of $m\mapsto Q_{b,\xi}^+(m)$ in $[-\mstar,\mstar]$. Moreover, $m_+(b,\xi)$ lies in $(-\mstar,T^{-1}(\xi))$ and is strictly increasing in~$b$.
\item[(b)] 
For $b = b_1(\xi)$, the function $m\mapsto Q_{b,\xi}^+(m)$ has exactly two minimizers,
$\mstar$ and a value $m_1(\xi) \in (-\mstar,T^{-1}(\xi))$.
\item[(c)] 
We have $b_2(\xi) = E_\xi'(-\mstar)$.
\item[(d)] 
The non-trivial minimizer in~(ii), $m_1(\xi)$, is the unique solution of
\begin{equation}
\label{3.28}
    E_\xi(m) + (\mstar - m)E_\xi'(m) = E_\xi(\mstar).
\end{equation}
Moreover, we have
\begin{equation}
\label{3.29}
b_1(\xi)=E_\xi'\bigl(m_1(\xi)\bigr).
\end{equation}
\item[(e)] 
As~$b$ tends to the boundaries of the interval~$(b_1(\xi),b_2(\xi))$, the unique minimizer in~(a) has the following limits
\begin{equation}
\lim_{b \downarrow b_2(\xi)} m_+(b,\xi) = -\mstar
\quad\text{and}\quad
\lim_{b \uparrow b_1(\xi)} m_+(b,\xi) =m_1(\xi),
\end{equation}
where~$m_1(\xi)$ is as in~(b). Both limits are uniform on compact subsets of~$(\xit,\infty)$.
\end{enumerate}
Now, part~(3) of the theorem follows from~(a) while the explicit formula~\eqref{2.13} for $b_2(\xi)$ for $\xi\geq \xit$ is readily derived from~(c).  
For, $\xi \leq \xit$, the critical curve $\xi\mapsto b_2(\xi)$ is given by the relation
$Q_{b,\xi}^+(\mstar) = Q_{b,\xi}^+(-\mstar)$, which gives also
the $\xi\le\xit$ part of~\eqref{2.13}.  Continuity of $b\mapsto m_+(b,\xi)$ along the portion of $b = b_2(\xi)$ for~$\xi>\xit$ is implied by~(e), while the jump discontinuity at $b = b_1(\xi)$ is a consequence of~(a) and~(e). This proves part~(4) of the theorem.
\end{proofsect}

\begin{proofsect}{Proof of part~(2)}
It remains to prove the continuity of~$b_1'(\xi)$, identify the asymptotic of~$b_1'$ as~$\xi\to\infty$ and establish the strict concavity of~$\xi\mapsto b_1(\xi)$.
First we will show that the non-trivial minimizer,~$m_1(\xi)$, is strictly increasing with~$\xi$. Indeed, we write \eqref{3.28} as~$F_\xi(m)=0$, where~$F_\xi(m)=E_\xi(\mstar)-E_\xi(m)-(\mstar-m)E_\xi'(m)$. Now,
\begin{equation}
\label{3.31}
\frac\partial{\partial\xi}F_\xi(m)=g(m)-g(\mstar)+(\mstar-m)g'(m),
\end{equation}
which is positive for all~$m\in[-\mstar,\mstar)$ by strict concavity of~$g$. Similarly,
\begin{equation}
\label{3.32}
\frac\partial{\partial m}F_\xi(m)=-E_\xi''(m)(\mstar-m),
\end{equation}
which at~$m=m_1(\xi)$ is negative because~$m_1$ lies in the convexity interval of~$E_\xi$, i.e., $m_1(\xi) \in (-\mstar,T^{-1}(\xi))$.
From (d) and implicit differentiation we obtain that $m_1'(\xi)>0$ for $\xi > \xit$.
By \eqref{3.29} we then have
\begin{equation} 
\label{dbdxi}
    b_1'(\xi) = -\frac{g(\mstar)-g(m_1)}{\mstar-m_1}
\end{equation}
which, invoking the strict concavity of~$g$ and the strict monotonicity of~$m_1$, implies that~$b_1'(\xi)>0$, i.e.,~$b_1$ is strictly convex on $(\xit,\infty)$.

To show the remaining items of~(2), it suffices to establish the limits
\begin{equation}
\label{3.34}
\lim_{\xi\downarrow\xit}m_1(\xi)=-\mstar
\quad\text{and}\quad
\lim_{\xi\to\infty}m_1(\xi)=\mstar.
\end{equation}
Indeed, using the former limit in \eqref{dbdxi} we get that~$b_1'(\xi)\to-g'(\mstar)$ as~$\xi\to\infty$ while the latter limit and~(c) above yield that~$b_1'(\xi)\to b_2'(\xit)$ as~$\xi\downarrow\xit$ which in light of the fact that~$b_1(\xi)=b_2(\xi)$ for~$\xi\le\xit$ implies the continuity of~$b_1'$. To prove the left limit in \eqref{3.34}, we just note that, by \eqref{3.28}, the slope of~$\oE_\xi$ at~$m=m_1(\xi)$ converges to zero as~$\xi\downarrow\xit$. Invoking the convexity/concavity picture, there are two points on the graph of~$m\mapsto\oE_{\xit}(m)$ where the slope is zero:~$\mstar$ and the absolute maximum of~$\oE_\xi$. The latter choice will never yield a minimizer of~$Q_{b,\xi}^+$ and so we must have~$m_1(\xi)\to\mstar$ as claimed. The right limit in \eqref{3.34} follows from the positivity of the quantity in \eqref{3.31}. Indeed, for each~$m\in[-\mstar,\mstar)$ we have~$F_\xi(m)>0$ once~$\xi$ is sufficiently large. Hence,~$m_1(\xi)$ must converge to the endpoint~$\mstar$ as~$\xi\to\infty$.
\end{proofsect}

\subsection{Remaining proofs}
\label{sec3.3}\noindent
Here we will prove Theorem~\ref{thm4b}, which
describes the phase diagrams for the ``ice'' boundary condition, and Theorem~\ref{thm5} which
characterizes the spin-sector of the distributions~$P_{W_L}^{\pm,c_L,h_L}$.

\smallskip
For the duration of the proof of Theorem~\ref{thm4b}, we will use the functions $E_\xi$ and~$\oE_\xi$ from \twoeqref{3.24}{3.25} with $\scrM = \scrM_{+,J}$ replaced by $\scrM = \scrM_{-,J}$. The main difference caused by this change is that the function~$m\mapsto\oE_\xi(m)$ may now have more complicated convexity properties. Some level of control is nevertheless possible:

\begin{lemma}
\label{lemma3.5}
There are at most two points inside~$[-\mstar,\mstar]$ where the second derivative of function $m\mapsto\oE_\xi(m)$ changes its sign.
\end{lemma}

\begin{proofsect}{Proof}
Consider again the function $T(m)=\scrM''(m)/g''(m)$ which characterizes~$\oE_\xi''(m)>0$ by $T(m)>\xi$. In the present cases, this function is given~by the expression
\begin{equation}
T(m) = \frac{\scrM''(m)}{g''(m)} = C(\mstar + m)^{-\frac{d+1}d}\bigl(m+\cot(\kappa/2)\bigr)^2
\end{equation}
where~$C=C(J)>0$ is a constant. Clearly,~$T$ starts off at plus infinity at $m=-\mstar$ and decreases for a while; the difference compared to the situation in Theorem~\ref{thm4} is that~$T$ now need not be monotone. Notwithstanding, taking the obvious extension of~$T$ to all~$m\ge-\mstar$, there exists a value~$\mT\in(-\mstar,\infty)$ such that~$T$ is decreasing for~$m<\mT$ while it is increasing for all~$m>\mT$. Now two possibilities have to be distinguished depending on whether~$\mT$ falls in or out of the interval~$[-\mstar,\mstar)$:
\settowidth{\leftmargini}{(1111)}
\begin{enumerate}
\item[(1)]
$\mT\ge\mstar$, in which case the equation~$T(m)=\xi$ has at most one solution for every~$\xi$ and~$m\mapsto\oE_\xi(m)$ is strictly concave on~$[-\mstar,T^{-1}(\xi))$ and strictly convex on~$(T^{-1}(\xi),\mstar]$. (The latter interval may be empty.)
\item[(2)]
$\mT<\mstar$, in which case the equation~$T(m)=\xi$ has two solutions for~$\xi\in(T(\mT),T(\mstar)]$. Then~$m\mapsto\oE_\xi(m)$ is strictly convex between these two solutions and concave otherwise. The values of~$\xi$ for which there is at most one solution to~$T(m)=\xi$ inside~$[-\mstar,\mstar]$ reduce to the cases in~(1). (This includes~$\xi=T(\mT)$.)
\end{enumerate}
We conclude that the type of convexity of~$m\mapsto\oE_\xi(m)$ changes at most twice inside the interval~$[-\mstar,\mstar]$, as we were to prove.
\end{proofsect}

The proof will be based on studying a few cases depending on the order of the control parameters~$\xi_1$ and~$\xi_2$ from \eqref{thexis}. The significance of these numbers for the problem at hand will become clear in the following lemma:

\begin{lemma}
\label{lemma3.6}
The derivatives $\oE_\xi'(\mstar)$ and $\oE_\xi''(\mstar)$ are strictly increasing functions of~$\xi$. In particular, for~$\xi_1$ and~$\xi_2$ as defined in \eqref{thexis}, we have
\begin{enumerate}
\item[(1)]
$\oE_\xi'(\mstar)<0$ if $\xi<\xi_1$ and $\oE_\xi'(\mstar)>0$ if $\xi>\xi_1$.
\item[(2)]
$\oE_\xi''(\mstar)<0$ if $\xi<\xi_2$ and $\oE_\xi''(\mstar)>0$ if $\xi>\xi_2$.
\end{enumerate}
\end{lemma}

\begin{proofsect}{Proof}
This follows by a straightforward calculation.
\end{proofsect}

Now we are ready to prove the properties of the phase diagram for minus boundary conditions:

\begin{proofsect}{Proof of Theorem~\ref{thm4b}}
Throughout the proof, we will regard the graph of the function~$m\mapsto\oE_\xi(m)$ as evolving dynamically---the role of the ``time'' in this evolution will be taken by~$\xi$.
We begin by noting that, in light of the strict concavity of function~$g$ from \eqref{g}, the value $\oE_\xi(m)$ is strictly decreasing in~$\xi$ for all~$m\in(-\mstar,\mstar)$. This allows us to define
\begin{equation}
\txit = \inf\bigl\{ \xi\ge0\colon \oE_\xi(m) < 0 \text{ for some } m\in(-\mstar,\mstar) \bigr\}.
\end{equation}
Now for~$\xi=0$ we have~$\oE_\xi(m)>0$ for all~$m\in(-\mstar,\mstar)$ while for~$\xi>\xi_1$, the minimum of~$\oE_\xi$ over~$(-\mstar,\mstar)$ will be strictly negative. Hence, we have~$0<\xit\le\xi_1$.

We will also adhere to the geometric interpretation of finding the mimizers of~$m\mapsto Q^-_{b,\xi}(m)$, cf proof of part~(1) of Theorem~\ref{thm4}. In particular, for each~$\xi>0$ we have two values~$\tb_1$ and~$\tb_2$ with~$\tb_2\le\tb_1$ such that the extremes $-\mstar$ and~$\mstar$ are the unique minimizers for~$b<\tb_2$ and~$b>\tb_1$, respectively, while none of these two are minimizers when~$\tb_2<b<\tb_1$. Here we recall that~$\tb_1$ is the minimal slope such that a straight line with this slope touches the graph of~$\oE_\xi$ at~$\mstar$ and at some other point, but it never gets above it, and similarly~$\tb_2$ is the maximal slope of a line that touches the graph of~$\oE_\xi$ at~$-\mstar$ and at some other point, but never gets above it.

As a consequence of the above definitions, we may already conclude that~(1) is true. (Indeed, for~$\xi\le\txit$ we have~$\oE_\xi(m)\ge0$ and so the two slopes~$\tb_1$ and~$\tb_2$ must be the same. For~$\xi>\txit$ there will be an~$m$ for which~$\oE_\xi(m)<0$ and so~$\tb_1\ne\tb_2$.)
The rest of the proof proceeds by considering two cases depending on the order of~$\xi_1$ and~$\xi_2$. We begin with the easier of the two,~$\xi_1\ge\xi_2$:

\medskip\noindent
\textsl{CASE~$\xi_1\ge\xi_2$:}
Here we claim that the situation is as in Theorem~\ref{thm4} and, in particular,~$\txit=\xi_1$. Indeed, consider a~$\xi>\xi_2$ and note that~$\oE_\xi''(\mstar)>0$ by Lemma~\ref{lemma3.6}. Since $\oE_\xi''(m)$ is negative near~$m=-\mstar$ and positive near~$m=\mstar$, it changes its sign an odd number of times. In light of Lemma~\ref{lemma3.5}, only one such change will occur and so~$[-\mstar,\mstar]$ splits into an interval of strict concavity and strict convexity of~$m\mapsto\oE_\xi(m)$. Now, if~$\txit$ is not equal~$\xi_1$, we may choose~$\xi$ between~$\txit$ and~$\xi_1$ so that~$\oE_\xi'(\mstar)<0$. This implies that $\oE_\xi(m)>0$ for all~$m<\mstar$ in the convexity region; in particular, at the dividing point between concave and convex behavior. But then a simple convexity argument $\oE_\xi(m)>0$  throughout the concavity region (except at~$-\mstar$). Thus~$\oE_\xi(m)>0$ for all~$m\in(-\mstar,\mstar)$ and so we have~$\xi\le\txit$. It follows that~$\txit=\xi_1$.

Invoking the convexity/concavity picture from the proof of Theorem~\ref{thm4} quickly finishes the argument. Indeed, we immediately have~(4) and, letting~$\txiu=\txit$, also the corresponding portion of~(5). It remains to establish the properties of~$\tb_1$ and~$\tb_2$---this will finish both~(2) and~(3a). To this end we note that~$\tb_1$ is determined by the slope of~$E_\xi$ at~$\mstar$, i.e., for~$\xi\ge\txit$,
\begin{equation}
\label{3.37}
\tb_1(\xi)=E_\xi'(\mstar).
\end{equation}
This yields the second line in \eqref{2.19}; the first line follows by taking the slope of~$E_\xi$ between~$-\mstar$ and~$\mstar$. As for~$\tb_2$, here we note that an analogue of the argument leading to \eqref{dbdxi} yields
\begin{equation}
\label{3.38}
\tb_2'(\xi)=-\frac{g(m_1)-g(-\mstar)}{m_1+\mstar},
\qquad\xi\ge\txit,
\end{equation}
where~$m_1=m_1(\xi)$ is the non-trivial minimizer at~$b=\tb_2(\xi)$. In this case the argument analogous to \twoeqref{3.31}{3.32} gives~$m_1'(\xi)<0$. The desired limiting values (and continuity) of~$\tb_2'$ follow by noting that~$m_1(\xi)\to\mstar$ as~$\xi\downarrow\txit$ and~$m_1(\xi)\to-\mstar$ as~$\xi\to\infty$.

\medskip\noindent
\textsl{CASE~$\xi_1<\xi_2$:}
Our first item of business is to show that~$\txit<\xi_1$. Consider the situation when~$\xi=\xi_1$ and~$m=\mstar$. By Lemma~\ref{lemma3.6} and continuity, the derivative~$\oE_{\xi_1}'(\mstar)$ vanishes, but, since we are assuming~$\xi_1<\xi_2$, the second derivative~$\oE_{\xi_1}''(\mstar)$ has not ``yet'' vanished, so it is still negative. The upshot is that~$\mstar$ is a local maximum for~$m\mapsto\oE_{\xi_1}(m)$. In particular, looking at~$m$ slightly less than~$\mstar$, we must encounter negative values of~$\oE_{\xi_1}$ and, eventually, a minimum of~$\oE_{\xi_1}$ in~$(-\mstar,\mstar)$. This implies that~$\txit<\xi_1$. 

Having shown that~$\txit<\xi_1<\xi_2$, we note that for~$\xi\in(\txit,\xi_2)$, the function~$m\mapsto\oE_\xi(m)$ changes from concave to convex to concave as~$m$ increases from~$-\mstar$ to~$\mstar$, while for $\xi\ge\xi_2$, exactly one change of convexity type occurs. Indeed,~$\oE_\xi$ is always concave near~$-\mstar$ and, when~$\xi<\xi_2$, it is also concave at~$\mstar$. Now, since~$\xi>\txit$, its minimum occurs somewhere in~$(-\mstar,\mstar)$. This implies an interval of convexity. But, by Lemma~\ref{lemma3.5}, the convexity type can change only at most twice and so this is all that we can have. For the cases~$\xi>\xi_2$ we just need to realize that~$\oE_\xi$ is now convex near~$m=\mstar$ and so only one change of convexity type can occur. A continuity argument shows that the borderline situation,~$\xi=\xi_2$, is just like~$\xi>\xi_2$.

The above shows that the cases~$\xi\ge\xi_2$ are exactly as for~$\xi_1\ge\xi_2$ (or, for that matter, Theorem~\ref{thm4}) while~$\xi<\txit$ is uninteresting by definition, so we can focus on~$\xi\in[\txit,\xi_2)$. Suppose first that~$\xi>\txit$ and let~$I_\xi$ denote the interval of strict convexity of~$\oE_\xi$. The geometrical minimization argument then shows that, at~$b=\tb_1$, there will be exactly two minimizers,~$\mstar$ and a value~$m_1(\xi)\in I_\xi$, while at~$b=\tb_2$, there will also be two minimizers,~$-\mstar$ and a value~$m_2(\xi)\in I_\xi$. For~$\tb_1<b<\tb_2$, there will be a unique minimizer~$m_-(b,\xi)$ which varies between~$m_2(\xi)$ and~$m_1(\xi)$. Since~$\oE_\xi$ is strictly convex in~$I_\xi$, the map
$b\mapsto m_-(b,\xi)$ is strictly increasing with limits~$m_1(\xi)$ as~$b\uparrow\tb_1(\xi)$ and~$m_2(\xi)$ as~$b\downarrow\tb_2(\xi)$. Both $m_1$ and~$m_2$ are inside~$(-\mstar,\mstar)$ so~$m_-$ undergoes a jump at both~$\tb_1$ and~$\tb_2$. Clearly,~$m_1(\xi)\ne m_2(\xi)$ for all~$\xi\in(\txit,\xi_2)$.

At~$\xi=\txit$, there will be an ``intermediate'' minimizer, but now there is only one. Indeed, the limits of~$m_1(\xi)$ and~$m_2(\xi)$ as~$\xi\downarrow\txit$ must be the same because otherwise, by the fact that $[m_1(\xi),m_2(\xi)]$ is a subinterval of the convexity interval~$I_\xi$, the function $\oE_{\txit}$ would vanish in a whole \emph{interval} of~$m$'s, which is impossible. Denoting the common limit by~$m_0$ we thus have three minimizers at~$\xi=\txit$; namely, $\pm\mstar$ and~$m_0$. This proves part~(4) and, letting~$\txiu=\xi_2$, also part~(5) of the theorem. As for the remaining parts, the strict concavity of~$\tb_1$ and the limits \eqref{derivs} are again consequences of formulas of the type \eqref{dbdxi} and \twoeqref{3.37}{3.38} and of the monotonicity properties of~$m_1$ and~$m_2$. The details are as for the previous cases, so we will omit them.
\end{proofsect}

\begin{proofsect}{Proof of Theorem~\ref{thm5}} As in~Part~I, the representation
\eqref{2.6b} is a simple consequence of the absence of salt-salt interaction as
formulated in Lemma~\ref{lemma3.2}. The fact that any subsequential (weak)
limit~$\rho^\pm$ of~$\rho_L^\pm$ has all of its mass concentrated on the
minimizers of~$Q^\pm_{b,\xi}$ is a consequence of Theorem~\ref{thm3} and the fact
that~$m$ can only take~$O(L)$ number of distinct values. Moreover, if the
minimizer is unique, which for the plus boundary conditions happens when~$b\ne
b_1(\xi),b_2(\xi)$, any subsequential limit is the Dirac mass at the unique
minimum (which is~$m_+(b,\xi)$ for the plus boundary conditions and $m_-(b,\xi)$ for the minus boundary conditions).
\end{proofsect}

\section*{Acknowledgments}
\noindent 
The research of K.S.A. was supported by the NSF under the 
grants DMS-0103790 and DMS-0405915.
The research of M.B. and L.C.~was supported by
the NSF grant DMS-0306167.

\end{document}